\documentclass[12pt]{article}
\usepackage[cp1251]{inputenc} 
\usepackage[english, russian]{babel} 
\usepackage{amsmath, amssymb, amsfonts, bm, array, graphics, eucal, cite}
\usepackage[dvips]{graphicx}
\usepackage[flushleft,small]{caption2} 

\makeatletter
\newcommand{\diag}{\rm \diag\, }

\renewcommand{\Re}{\mathop{\rm Re\,}}
\renewcommand{\Im}{\mathop{\rm Im\,}}
\makeatother

\begin{document}
\thispagestyle{empty}

\pagestyle{myheadings}
\textwidth=160 mm
\textheight=240 mm
\baselineskip=24pt
\voffset=-10mm
\hoffset=-10mm
\makeatletter
\renewcommand{\abstractname}{\ }
\renewcommand{\refname}{\begin{center}\normalsize Cписок литературы
 \end{center}}

\large
\begin{center}\bf
Mass velocity of Bose-gas in the problem about isothermal sliding
\end{center}

\begin{center}
{\bf E. A. Bedrikova\footnote{e-mail: bedrikova@mail.ru} and
A. V. Latyshev\footnote{e-mail: avlatyshev@mail.ru}}
\end{center}

\begin{center}
{\it Faculty of Physics and Mathematics,\\ Moscow State Regional
University, 105005,\\ Moscow, Radio str., 10--A}
\end{center}\medskip

\begin{abstract}
Distribution of mass velocity of quantum Bose-gas in half-space is
received. Far from half-space border the gradient of mass
velocity is set. The mass velocity of Bose-gas directly at a wall is found
also. All results are received on basis of the analytical solution
of Kramer's problem. The analysis of dependence of velocity
coefficients from parametre quantity is carried out. The parametre
representing the relation of chemical potential to  product of
Boltzmann constant on absolute temperature. Graphic comparison of
coefficients of velocity Bose- and Fermi-gases is resulted.

Получено распределение массовой скорости квантового бозе--газа в
полупространстве. Вдали от границы полупространства задан
градиент массовой скорости. Отыскивается также скорость
Бозе--газа непосредственно у стенки. Все результаты получены на
основе аналитического решения задачи Крамерса.
Проведен анализ зависимости коэффициента скорости от величины
параметра, представляющего собой отношение химического потенциала к
произведению постоянной Больцмана на абсолютную температуру.
Приводится графическое сравнение коэффициентов скорости Бозе-- и
Ферми--газов.
\end{abstract}

\begin{center}
\bf  Введение
\end{center}

Кинетическое уравнение Больцмана было обобщено на случай
квантовых газов в работе \cite{Uling}. Систематическое решение
граничных задач для квантовых газов было проведено в работах
\cite{LatTMF01}--\cite{LatTMF03}. При этом использовались
кинетические уравнения с модельным интегралом столкновений \cite{Cerc73}
типа БГК (Бхатнагар, Гросс, Крук)
$$
M[f]=\dfrac{f_{eq}-f}{\tau},
$$
где $\tau=1/\nu$ -- характерное время между двумя
последовательными столкновениями частиц, $\nu$ -- эффективная
частота столкновений частиц, $f_{eq}$ -- локально равновесная
функция распределения Ферми---Дирака или Бозе---Эйнштейна
$$
f_{eq}^F=\dfrac{1}{1+\exp\dfrac{E-\mu}{k_BT}}
$$
или
$$
f_{eq}^B=\dfrac{1}{-1+\exp\dfrac{E-\mu}{k_BT}}.
$$

Здесь $E=\dfrac{m(\mathbf{v}-\mathbf{u})^2}{2}$ -- кинетическая
энергия частиц газа, $\mathbf{u}$ -- массовая скорость газа,
$\mathbf{v}$ -- скорость частиц газа, $m$ -- масса частицы,
$k_B$ -- постоянная Больцмана, $T$ -- температура газа, $\mu$ --
химический потенциал газовых молекул (частиц).

В работах \cite{LatJTF09}--\cite{LatFNT10} было продолжено
аналитическое решение
граничных задач для квантовых газов. Так, в работе \cite{LatJTF09} было
получено решение задачи о скачке химического потенциала при
испарении квантового Ферми--газа, в \cite{LatFNT08} получено решение
задачи Крамерса для Ферми--газа с аккомодационными граничными
условиями, в \cite{LatIVUS09} и \cite{LatFNT10} получены решения задачи
Крамерса соответственно для квантовых Ферми-- и Бозе--газов с частотой
столкновений частиц, пропорциональной модулю их скорости.

Работа \cite{Lubimova} посвящена точному решению задачи
Максвелла о тепловом скольжении Ферми--газа.

В работах \cite{LatTMF08}--\cite{LatTMF10b} изучались проблемы
температурного скачка
(сопротивление Капицы) в вырожденных квантовых Бозе--газах.

Настоящая работа является продолжением работы \cite{LatIVUS02}. В
\cite{LatIVUS02} была решена полупространственная задача Крамерса для
квантовых Бозе--газов: найдена скорость изотермического
скольжения вдоль плоской поверхности и построена функция
распределения летящих к стенке молекул непосредственно у стенки.

В настоящей работе для задачи Крамерса построено распределение
массовой скорости Бозе--газа в полупространстве и находится ее
значение непосредственно у стенки. Для этого выводится формула
факторизации дисперсионной функции. Проводится сравнение
коэффициентов массовой скорости Бозе-- и Ферми--газов.

\begin{center}
  \bf 1. Постановка задачи и основные уравнения
\end{center}

Пусть газ занимает полупространство $x>0$
над плоской стенкой и движения вдоль оси $y$ с массовой скоростью
$u_y(x)$. Вдали от поверхности имеется градиент массовой скорости
газа, т.е. профиль скорости представим в виде
$u_y(x)=u_{sl}+g_vx,\; x\to +\infty$. Величина $u_{sl}$ называется
скоростью изотермического скольжения,
$g_v=\Big(\dfrac{du}{dx}\Big)_{x=\infty}$.
При малых градиентах $g_v$ скорость изотермического скольжения
пропорциональна величине градиента:
$u_{sl}=K_vlg_v$.
Здесь $K_v$ -- коэффициент изотермического скольжения,
$l$ -- средняя длина свободного пробега частиц.

Величина $K_v$ определяется кинетическими процессами вблизи
поверхности. Для ее определения необходимо решить кинетическое уравнение
в так называемом слое Кнудсена, т.е. в слое газа, примыкающего к поверхности,
толщиной порядка длины свободного пробега $l$.

В работе \cite{LatIVUS02} выведено линеаризованное кинетическое уравнение
для Бозе--газа
$$
\mu\dfrac{\partial h}{\partial
x_1}+h(x_1,\mu)=\int\limits_{-\infty}^{\infty}K(\mu',\alpha)
h(x_1,\mu')d\mu',\quad x_1>0, \quad \alpha<0.
\eqno{(1.1)}
$$

Функция $h(x_1,\mu)$ связана с функцией распределения
$f(x,\mathbf{v})$ соотношением
$$
f=f_0(C)+C_yg(C)h(x_1,C_x).
$$

В этих соотношениях $x_1$ -- безразмерная координата, связанная
с размерной соотношением $x_1=x/l_T$, где $l_T$ -- тепловая
длина свободного пробега частиц Бозе--газа, $l_T=v_T\tau$, $v_T=
1/\sqrt{\beta}$, $\beta=m/(2k_BT)$, $v_T$ -- тепловая скорость
частиц, $K(\mu,\alpha)$ -- ядро уравнения,
$$
K(\mu,\alpha)=\dfrac{\ln(1-e^{\alpha-\mu^2})}{2l_0(\alpha)},\qquad
l_0(\alpha)=\int\limits_{0}^{\infty}\ln(1-e^{\alpha-\tau^2})d\tau,
$$
$\alpha=\mu_0/(k_BT)$, $\mu_0$ -- химический потенциал молекул
Бозе--газа,  $C=\sqrt{\beta}v$ -- модуль безразмерной скорости
частиц газа, $\mu=C_x$,
$$
f_0(C)=\dfrac{1}{-1+e^{C^2-\alpha}},\qquad
g(C)=\dfrac{\partial f_0}{\partial \alpha}=\dfrac{e^{C^2-\alpha}}
{(-1+e^{C^2-\alpha})^2}.
$$

Правая часть уравнения (1.1) есть удвоенная безразмерная
массовая скорость Бозе--газа,
$$
U(x_1,\alpha)=\dfrac{1}{2}\int\limits_{-\infty}^{\infty}K(\mu,\alpha)
h(x_1,\mu)d\mu.
\eqno{(1.2)}
$$

В случае диффузного отражения Бозе--частиц от стенки граничные
условия принимают вид:
$$
h(0,\mu)=0, \qquad \mu>0,
\eqno{(1.3)}
$$
$$
h(x_1,\mu)=h_{as}(x_1,\mu)+o(1),\qquad x_1\to +\infty,
\eqno{(1.4)}
$$
где
$$
h_{as}(x_1,\mu)=2U_{sl}+2G_v(x_1-\mu)
$$
-- распределение Чепмена---Энскога, $G_v$ -- градиент
безразмерной массовой скорости, заданный вдали от стенки,
$$
G_v=\left(\dfrac{dU(x_1,\alpha)}{dx_1}\right)_{x_1=+\infty}.
$$

Разделение переменных согласно подстановке
$$
h_\eta(x_1,\mu)=\exp\Big(-\dfrac{x_1}{\eta}\Big)\Phi(\eta,\mu)
$$
сводит уравнение (1.1) к характеристическому уравнению
$$
(\eta-\mu)\Phi(\eta,\mu)=\eta,
\eqno{(1.5)}
$$
в котором собственная функция $\Phi(\eta,\mu)$ нормирована
условием
$$
\int\limits_{-\infty}^{\infty}K(\mu,\alpha)\Phi(\eta,\mu)d\mu=1,\quad
\forall \alpha<0, \quad \forall \eta\in(-\infty,+\infty).
\eqno{(1.6)}
$$

Из уравнений (1.5) и (1.6) находим собственные функции,
отвечающие непрерывному спектру \cite{Vladimirov}
$$
\Phi(\eta,\mu)=\eta P\dfrac{1}{\eta-\mu}+\dfrac{\lambda(\eta,\alpha)}
{K(\eta,\alpha)}\delta(\eta-\mu).
$$
Здесь  символ $Px^{-1}$ означает распределение -- главное
значение интеграла при интегрировании выражения $x^{-1}$,
$\delta(x)$ -- дельта--функция Дирака, $\lambda(z,\alpha)$ --
дисперсионная функция,
$$
\lambda(z,\alpha)=1+z\int\limits_{-\infty}^{\infty}\dfrac{K(\mu,\alpha)
d\mu}{\mu-z}.
$$

В \cite{LatIVUS02} показано, что граничная задача (1.1), (1.3) и (1.4)
имеет решение в виде суммы распределения Чепмена---Энскога и
интеграла по непрерывному спектру
$$
h(x_1,\mu)=2U_{sl}(\alpha)+2G_v(x_1-\mu)+\int\limits_{0}^{\infty}
\exp\Big(-\dfrac{x_1}{\eta}\Big)\Phi(\eta,\mu)a(\eta)d\eta.
\eqno{(1.7)}
$$

В разложении (1.7) безразмерная скорость скольжения равна:
$$
U_{sl}(\alpha)=V_1(\alpha)G_v,
\eqno{(1.8)}
$$
а коэффициент непрерывного спектра определяется равенством
$$
a(\eta)=2G_v\dfrac{\sin \xi(\eta,\alpha)}{\pi \eta
X(\eta,\alpha)}.
\eqno{(1.9)}
$$

В равенствах (1.8) и (1.9)
$$
V_1(\alpha)=-\dfrac{1}{\pi}\int\limits_{0}^{\infty}\xi(\mu,\alpha)d\mu,
\qquad 
\xi(\mu,\alpha)=\theta(\mu,\alpha)-\pi,$$$$
\theta(\mu,\alpha)=\arg\lambda^+(\mu,\alpha)=
\arcctg\dfrac{\Re
\lambda^+(\mu,\alpha)}{\Im\lambda^+(\mu,\alpha)},
$$
$$
\lambda^{\pm}(\mu,\alpha)=\lambda(\mu,\alpha)\pm i\pi \mu
K(\mu,\alpha).
$$

Функция $X(z,\alpha)$ является решением задачи Римана
$$
\dfrac{X^+(\mu,\alpha)}{X^-(\mu,\alpha)}=\dfrac{\lambda^+(\mu,\alpha)}
{\lambda^-(\mu,\alpha)}, \qquad \mu>0,
$$
и имеет вид
$$
X(z,\alpha)=\dfrac{1}{z}\exp V(z,\alpha), \qquad
V(z,\alpha)=\dfrac{1}{\pi}\int\limits_{0}^{\infty}
\dfrac{\xi(\tau,\alpha)d\tau}{\tau-z}.
$$

\begin{center}
  \bf 2. Приведение формулы скорости скольжения к размерному
  виду
\end{center}

Формулу (1.8) для безразмерной скорости скольжения приведем к
размерному виду. Для этого понадобится коэффициент
кинематической вязкости
$$
\eta=-\dfrac{P_{xy}}{g_v},
\eqno{(2.1)}
$$
где
$$
P_{xy}=m\int fv_xv_yd\Omega,\qquad d\Omega=\dfrac{(2s+1)m^3d^3v}
{(2\pi\hbar)^3},
$$
$g_v=\nu G_v$, $s$ -- спин Бозе--частиц, $\hbar$ -- постоянная Планка.

Из (2.1) получаем, что кинематическая вязкость равна
$$
\eta=-\dfrac{m^4(2s+1)}{\nu G_v(2\pi\hbar)^3
(\sqrt{\beta})^5}\int f C_xC_yd^3C.
\eqno{(2.2)}
$$
Подставляя в (2.2) вместо $f$ ее выражение через $h(x_1,\mu)$,
имеем:
$$
\eta=-\dfrac{m^4(2s+1)}{\nu G_v(2\pi\hbar)^3
(\sqrt{\beta})^5}\int C_xC_y^2 g(C)h(x_1,\mu)d^3C.
\eqno{(2.3)}
$$
Устремляя в (2.3) $x_1$ к бесконечности, заменим здесь
$h(x_1,\mu)$ на $h_{as}(x_1,\mu)$. Получаем, что
$$
\eta=-\dfrac{2m^4(2s+1)}{\nu (2\pi\hbar)^3(\sqrt{\beta})^5}\int
C_x^2C_y^2g(C)d^3C=-\dfrac{10(2s+1)\pi m^4}{\nu(2\pi\hbar)^3
(\sqrt{\beta})^5}l_2(\alpha),
\eqno{(2.4)}
$$
где
$$
l_2(\alpha)=\int\limits_{0}^{\infty}\tau^2\ln(1-e^{\alpha-\tau^2})d\tau.
$$

Теперь нам понадобится числовая плотность частиц (концентрация)
в равновесном состоянии:
$$
N=\int f_0(C)d\Omega.
\eqno{(2.5)}
$$
На основании (2.5) получаем:
$$
N=\dfrac{4\pi
(2s+1)m^3}{(2\pi\hbar)^3(\sqrt{\beta})^3}\int\limits_{0}^{\infty}
C^2f_0(C)dC=-\dfrac{2\pi(2s+1)m^3}{(2\pi\hbar)^3(\sqrt{\beta})^3}
l_0(\alpha),
\eqno{(2.6)}
$$
где
$$
l_0(\alpha)=\int\limits_{0}^{\infty}\ln(1-e^{\alpha-\tau^2})d\tau.
$$
На основании (2.4) и (2.6) находим, что
$$
\dfrac{\eta}{N}=\dfrac{5ml_2(\alpha)}{\nu \beta l_0(\alpha)},
$$
откуда имеем:
$$
\eta=\dfrac{\rho}{\nu \beta}\dfrac{l_2(\alpha)}{l_0(\alpha)}.
\eqno{(2.7)}
$$

На основании (1.8) находим размерную скорость скольжения:
$
u_{sl}(\alpha)=V_1(\alpha)l_Tg_v,
$
или
$$
u_{sl}(\alpha)=\dfrac{V_1(\alpha)}{\nu \sqrt{\beta}l}lg_v.
\eqno{(2.8)}
$$
Среднюю длину свободного пробега газовых частиц выберем согласно
Черчиньяни \cite{Cerc73}: $l=\eta \sqrt{\pi \beta}/\rho$.
Подставим в это выражение вязкость согласно (2.7). Получаем, что
$$
l=\dfrac{\sqrt{\pi}}{\nu
\sqrt{\beta}}\cdot\dfrac{l_2(\alpha)}{l_0(\alpha)},\quad\text{или}\quad
l=l_T\dfrac{\sqrt{\pi}l_2(\alpha)}{l_0(\alpha)}.
$$
Следовательно, скорость изотермического скольжения Бозе--газа
вдоль плоской поверхности равна:
$$
u_{sl}(\alpha)=K_v(\alpha)lg_v,
\eqno{(2.9)}
$$
где
$$
K_v(\alpha)=\dfrac{V_1(\alpha)l_0(\alpha)}{\sqrt{\pi}l_2(\alpha)}
$$
-- коэффициент изотермического скольжения Бозе--газа вдоль
плоской поверхности.

\begin{center}
  \bf 3. Скорость Бозе--газа в полупространстве
\end{center}

Для нахождения массовой скорости Бозе--газа в полупространстве
воспользуемся формулой (1.2). Подставим в (1.2) разложение
(1.7). Воспользовавшись нормировкой (1.6), получаем
распределение безразмерной массовой скорости Бозе--газа в
полупространстве:
$$
U(x_1,\alpha)=U_{sl}(\alpha)+G_vx_1+\dfrac{1}{2}\int\limits_{0}^{\infty}
\exp\Big(-\dfrac{x_1}{\eta}\Big)a(\eta)d\eta.
\eqno{(3.1)}
$$

Воспользовавшись формулами (1.8) и (1.9) на основании (3.1)
получаем, что
$$
U(x_1,\alpha)=C_v(x_1,\alpha)G_v.
\eqno{(3.2)}
$$
Здесь
$$
C_v(x_1,\alpha)=V_1(\alpha)+x_1+\dfrac{1}{\pi}\int\limits_{0}^{\infty}
\exp\Big(-\dfrac{x_1}{\eta}\Big)\dfrac{\sin \xi(\eta,\alpha)d\eta}
{\eta X(\eta,\alpha)}.
\eqno{(3.3)}
$$

На основании (3.2) для размерной скорости Бозе--газа в
полупространстве получаем:
$$
u_y(x_1,\alpha)=C_v(x_1,\alpha)l_Tg_v,
$$
или, рассуждая, как и ранее, получаем
$$
u_y(x_1,\alpha)=K_v^*(x_1,\alpha)lg_v,
\eqno{(3.4)}
$$
где величина
$$
K_v^*(x_1,\alpha)\dfrac{C_v(x_1,\alpha)l_0(\alpha)}{\sqrt{\pi}
l_2(\alpha)}
$$
-- является коэффициентом распределения массовой скорости
Бозе--газа в полупространстве.

\begin{center}
  \bf 4. Скорость Бозе--газа непосредственно у стенки
\end{center}

Полагая $x_1=0$ в формуле (3.4), получаем:
$$
u_y(0,\alpha)=K_v^*(0,\alpha) lg_v,
\eqno{(4.1)}
$$
где
$$
K_v^*(0,\alpha)=\dfrac{C_v(0,\alpha)l_0(\alpha)}{\sqrt{\pi}l_2(\alpha)},
$$
$$
C_v(0,\alpha)=V_1(\alpha)+\dfrac{1}{\pi}\int\limits_{0}^{\infty}
\dfrac{\sin \xi(\eta,\alpha)d\eta}{\eta X(\eta,\alpha)}.
\eqno{(4.2)}
$$

Покажем, что интеграл из (4.2) можно вычислить аналитически. Это
означает, что скорость Бозе--газа (4.1) непосредственно у стенки
вычисляется без квадратур. Для этого потребуется интегральное
представление
$$
\dfrac{1}{X(z,\alpha)}=z-V_1(\alpha)-\dfrac{1}{\pi}\int\limits_{0}^{\infty}
\dfrac{\sin \xi(\eta,\alpha)d\eta}{X(\eta,\alpha)(\eta-z)}.
\eqno{(4.3)}
$$
Интегральное представление (4.3) выводится с помощью контурного
интегрирования так же, как и соответствующее интегральное
представление для классических газов \cite{LatMono08}.

На основании (4.3) для $z=\mu<0$ получаем:
$$
\dfrac{1}{\pi}\int\limits_{0}^{\infty}\dfrac{\sin \xi(\eta,\alpha)d\eta}
{X(\eta,\alpha)(\eta-\mu)}=\mu-V_1(\alpha)-\dfrac{1}{X(\mu,\alpha)}.
\eqno{(4.4)}
$$
Устремляя $\mu\to 0$ в (4.4), имеем:
$$
\dfrac{1}{\pi}\int\limits_{0}^{\infty}\dfrac{\sin \xi(\eta,\alpha)d\eta}
{X(\eta,\alpha)\eta}=-V_1(\alpha)-\dfrac{1}{X(-0,\alpha)}.
\eqno{(4.5)}
$$

Функция $X(\eta,\alpha)$ при $\eta\in (-\infty,+\infty)$
разрывна в нуле (см. \cite{LatMono08}). Для вычисления величины
$X(-0,\alpha)$ воспользуемся формулой факторизации дисперсионной
функции. Для ее вывода понадобится разложение дисперсионной
функции при $z\to \infty$ в асимптотический ряд
$$
\lambda(z,\alpha)=-\dfrac{\lambda_2(\alpha)}{z^2}-
\dfrac{\lambda_4(\alpha)}{z^4}-\cdots, \qquad z\to \infty.
\eqno{(4.6)}
$$
Здесь
$$
\lambda_{2n}(\alpha)=\dfrac{l_{2n}(\alpha)}{l_0(\alpha)},\qquad
n=1,2,\cdots.
$$

Рассуждая как и в \cite{LatMono08}, на основании (4.6) выводим формулу
факторизации дисперсионной функции
$$
\lambda(z,\alpha)=\dfrac{l_{2}(\alpha)}{l_0(\alpha)}
X(z,\alpha)X(-z,\alpha).
\eqno{(4.7)}
$$
Из формулы (4.7) получаем:
$$
X^2(0,\alpha)=\dfrac{l_{2}(\alpha)}{l_0(\alpha)}.
\eqno{(4.8)}
$$
Выбирая знак величины $X(0,\alpha)$ по непрерывности, на
основании (4.8) находим:
$$
X(-0,\alpha)=-\sqrt{\dfrac{l_0(\alpha)}{l_2(\alpha)}}.
\eqno{(4.9)}
$$
Возвращаясь к (4.2), получаем:
$$
C_v(0,\alpha)=-\dfrac{1}{X(-0,\alpha)}=
\sqrt{\dfrac{l_2(\alpha)}{l_0(\alpha)}}.
\eqno{(4.10)}
$$
Устремляя $\alpha\to -\infty$ в (4.10), получаем классический
результат Черчиньяни \cite{Cerc73}:
$$
C_v(0,-\infty)=\dfrac{1}{\sqrt{2}}.
$$

Таким образом, скорость Бозе--газа непосредственно у стенки
равна:
$$
u_y(0,\alpha)=\sqrt{\dfrac{l_0(\alpha)}{l_2(\alpha)}}lg_v.
\eqno{(4.11)}
$$

На рис. 1 изобразим зависимость коэффициента изотермического скольжения от
величины $\alpha$ -- величины отношения химического потенциала к
произведению постоянной Больцмана на абсолютную температуру.
При отрицательных $\alpha$
график выходит на свою асимптотику
уже при $\alpha\leqslant -4$.

Таким образом, анализ приведенного графика показывает, что переход между
вырожденным квантовым
и классическим случаем происходит на относительно узком интервале значений
$\alpha$. При $\alpha\leqslant -4$
величина коэффициента 
практически совпадает
с асимптотическим значением, соответствующим классическому случаю.

\begin{figure}[h]
\begin{center}
\includegraphics[width=14cm,height=10.5cm]{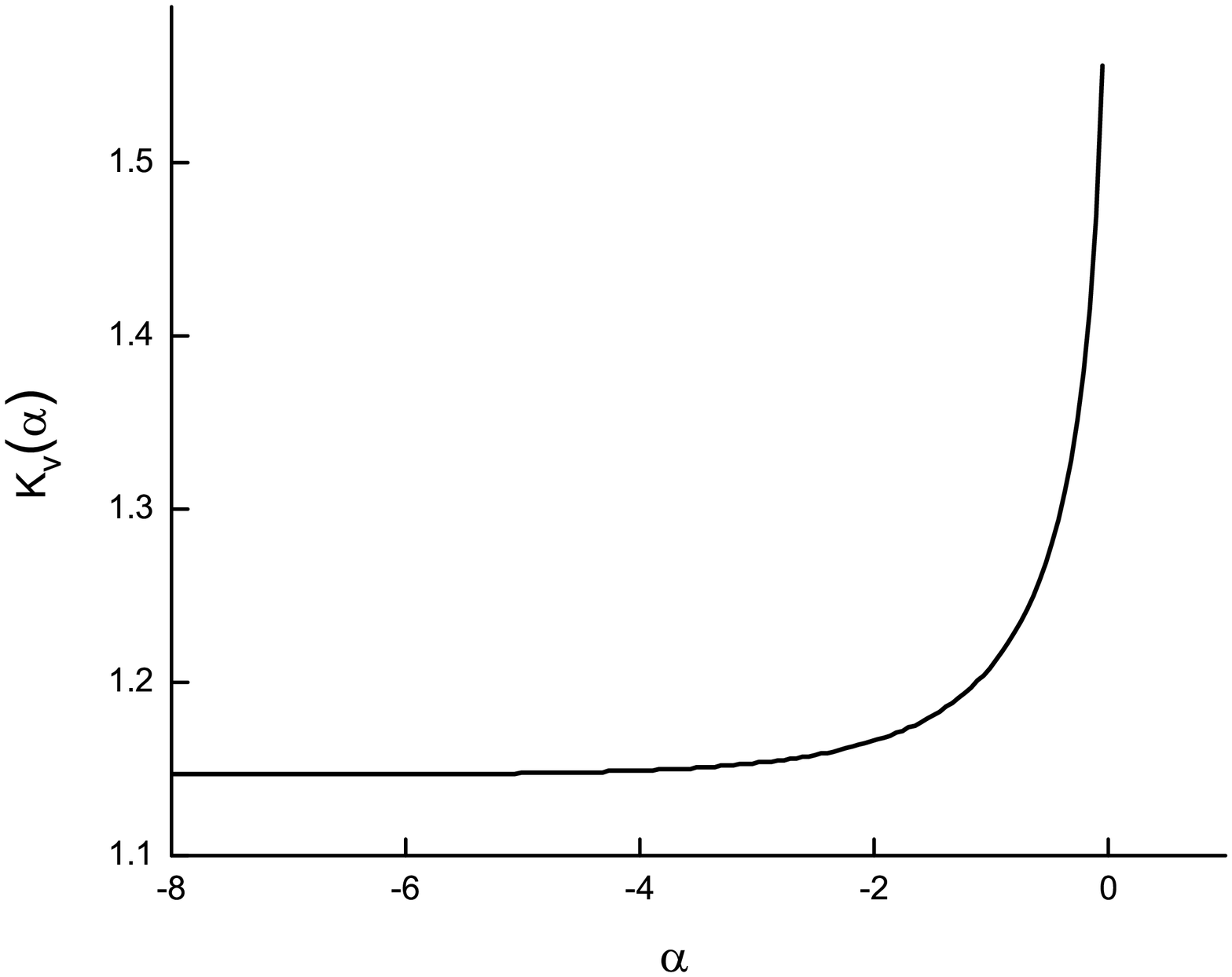}
\end{center}
\begin{center}
\caption{Зависимость коэффициента изотермического скольжения
Бозе--газа от приведенного химического потенциала $\alpha$.}
\end{center}
\end{figure}

\begin{figure}[h]
\begin{center}
\includegraphics[width=14cm,height=8.5cm]{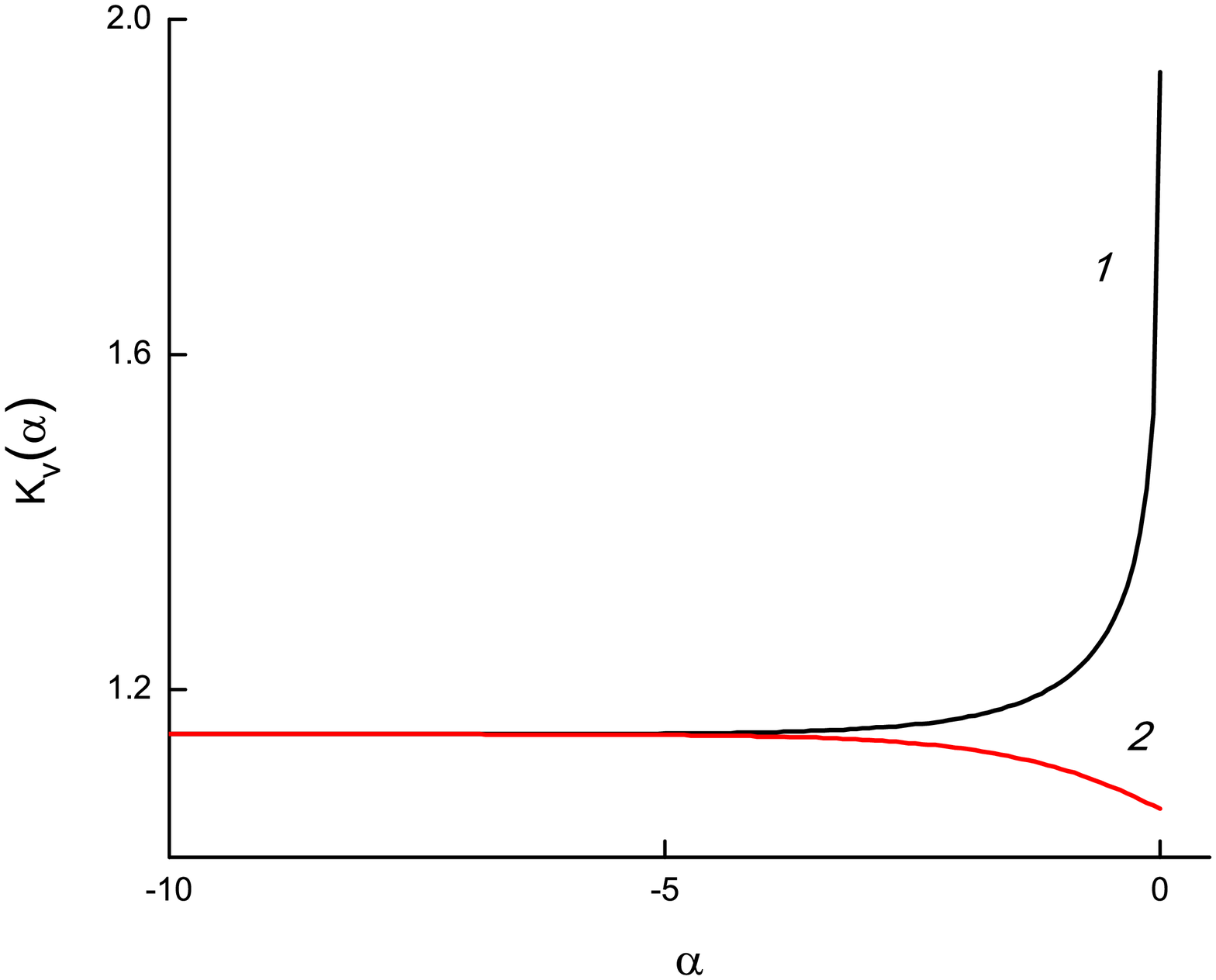}
\end{center}
\begin{center}
\caption{Зависимость коэффициентов изотермического скольжения
Бозе--газа  (кривая $1$) и Ферми--газа (кривая $2$) от приведенного
химического потенциала $\alpha$.}
\end{center}
\end{figure}

\begin{figure}[h]
\begin{center}
\includegraphics[width=14cm,height=8.5cm]{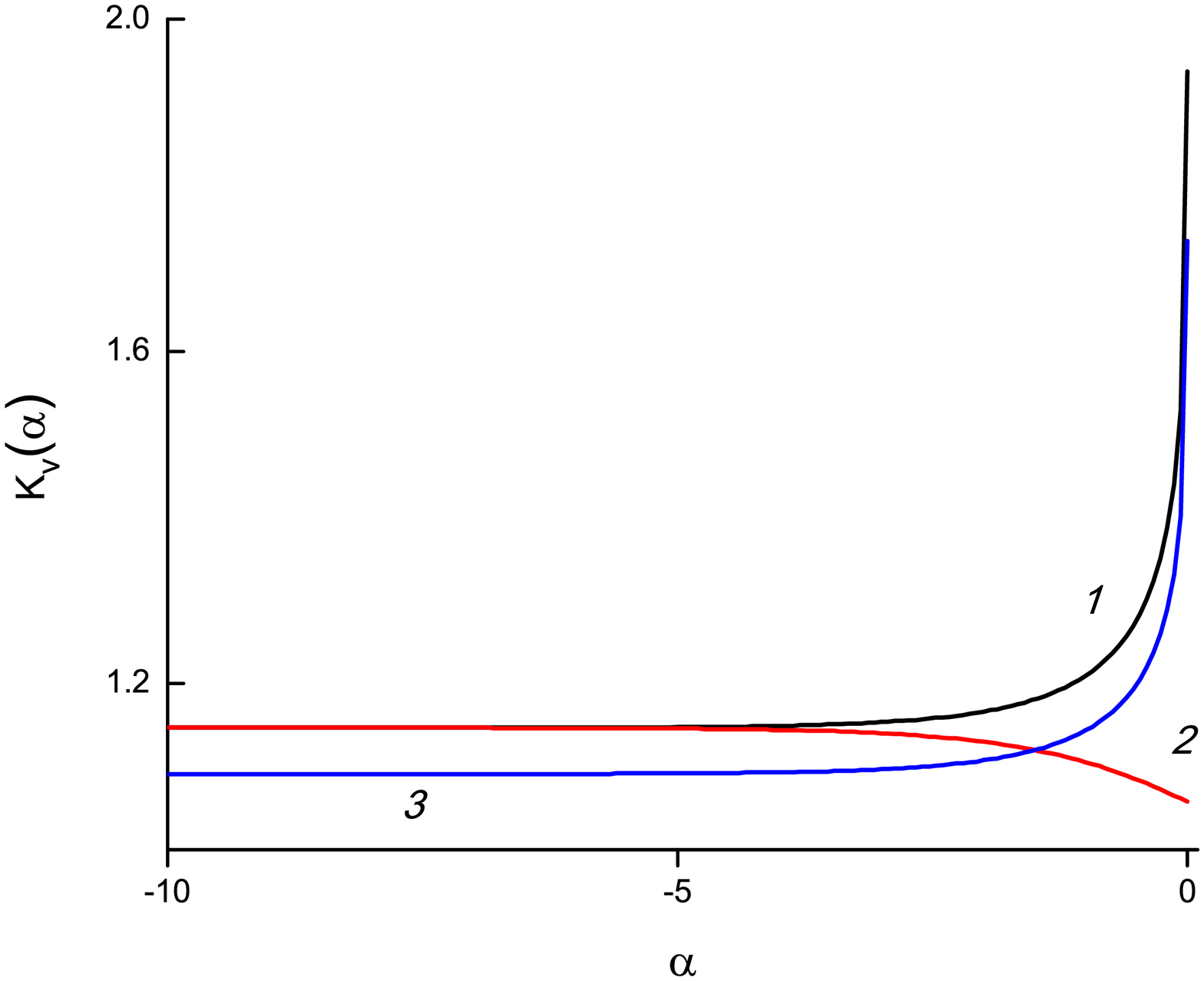}
\end{center}
\begin{center}
\caption{Зависимость коэффициентов изотермического скольжения
Бозе--газа  (кривая $1$), Ферми--газа (кривая $2$) и Бозе--газа с
частотой столкновений, пропорциональной модулю скорости (кривая $3$),
от приведенного химического потенциала $\alpha$.}
\end{center}
\end{figure}

\begin{figure}[h]
\begin{center}
\includegraphics[width=14cm,height=8.5cm]{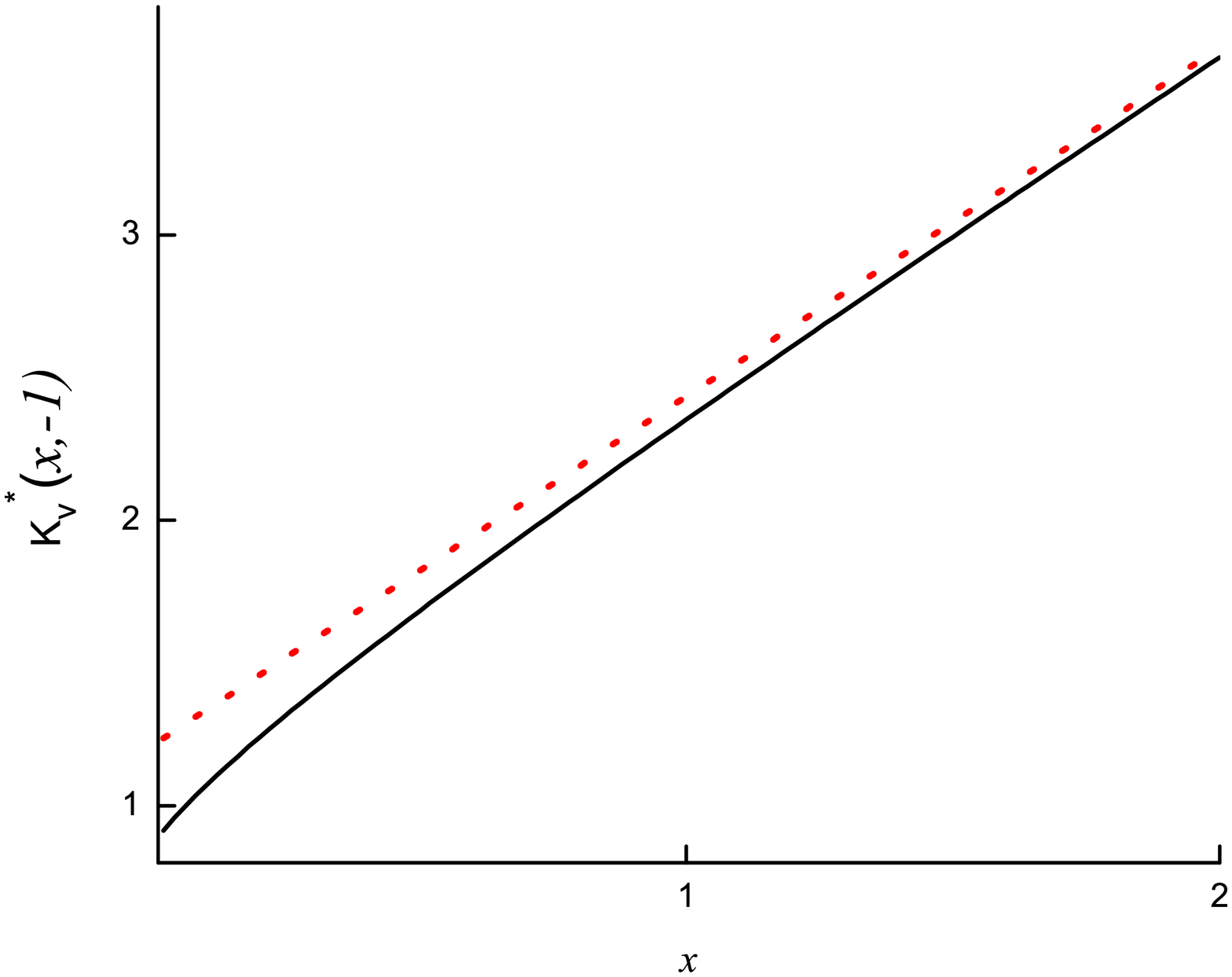}
\end{center}
\begin{center}
\caption{Профиль массовой скорости Бозе--газа в полупространстве.
Значение приведенного химического потенциала $\alpha=-1$.}
\end{center}
\end{figure}

\begin{figure}[h]
\begin{center}
\includegraphics[width=14cm,height=8.5cm]{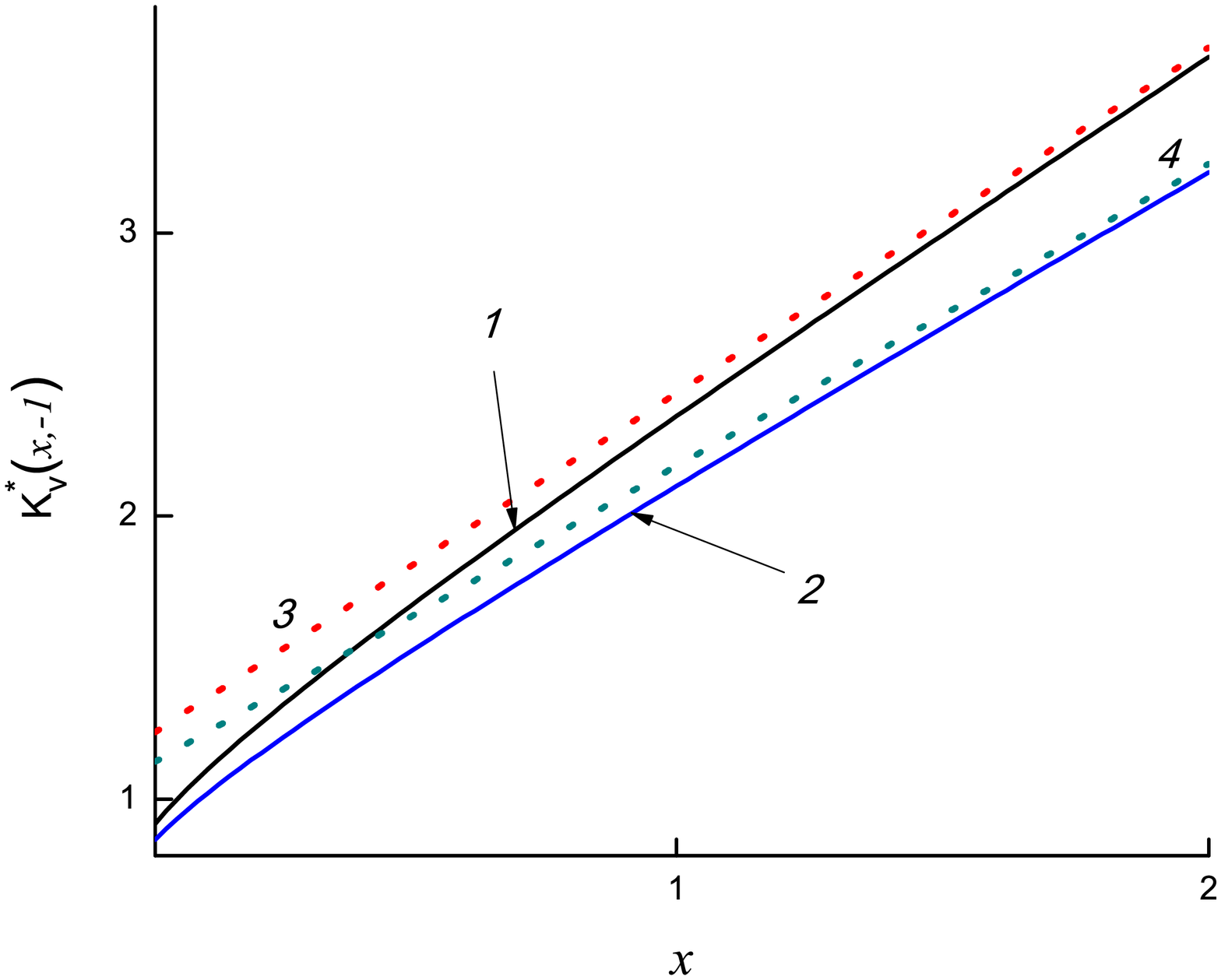}
\end{center}
\begin{center}
\caption{Профили массовой скорости Бозе--газа (кривая $1$, ее асимптотика --
кривая $3$) и
Ферми--газа (кривая $2$, ее асимптотика -- кривая $4$) в полупространстве.
Значение приведенного химического потенциала $\alpha=-1$.}
\end{center}
\end{figure}

\begin{figure}[h]
\begin{center}
\includegraphics[width=14cm,height=12.5cm]{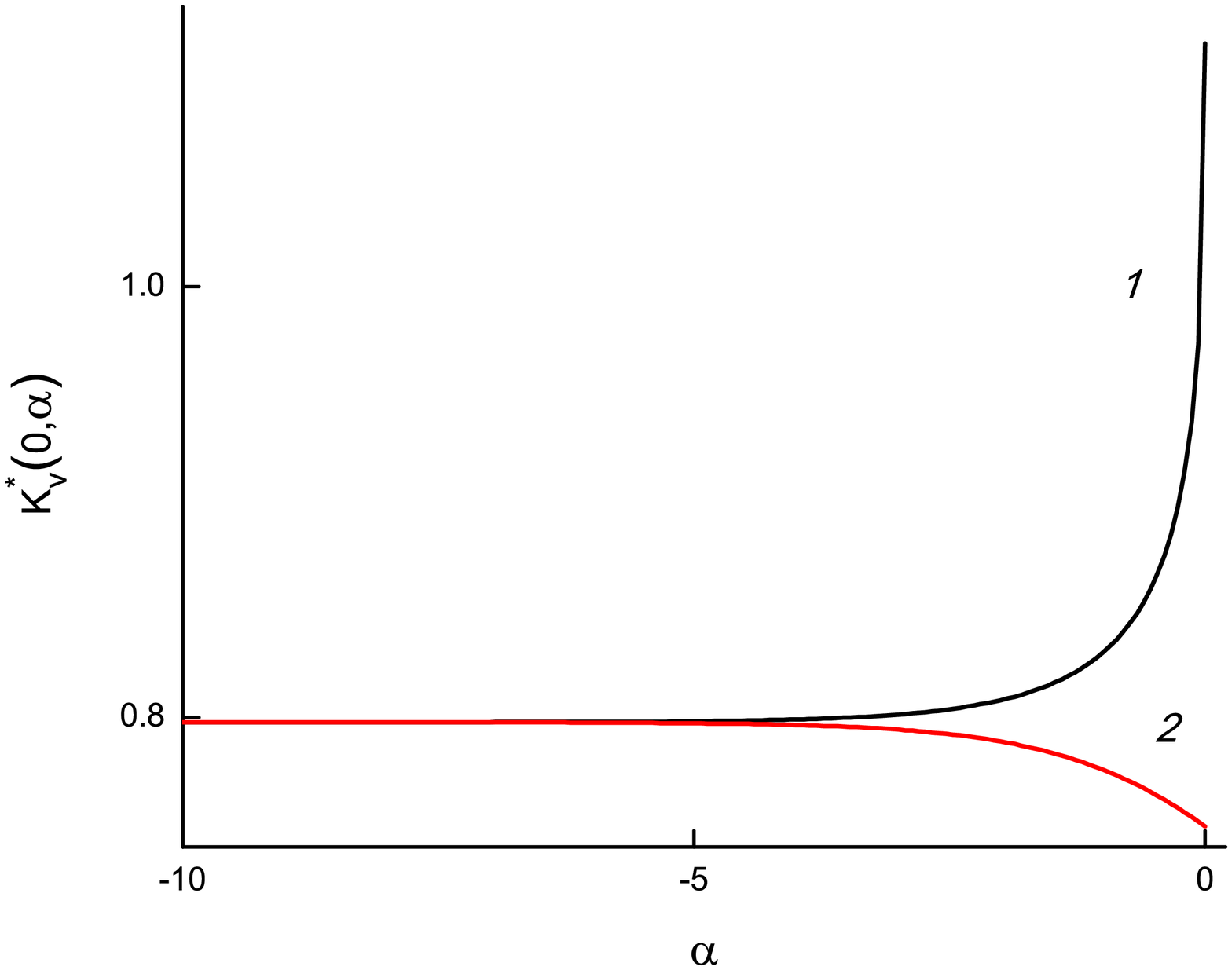}
\end{center}
\begin{center}
\caption{Зависимость коэффициентов массовой скорости Бозе--газа (кривая $1$)
и Ферми--газа (кривая $2$)
непосредственно у стенки от приведенного химического потенциала $\alpha$.}
\end{center}
\end{figure}
\clearpage
\begin{center}
5. ЗАКЛЮЧЕНИЕ
\end{center}

В настоящей работе приводится к размерному виду формула скорости
изотермического скольжения для квантового Бозе--газа. Приводится
графическое
сравнение коэффициентов скольжения Бозе-- и Ферми--газов.
Выведена формула профиля массовой скорости Бозе--газа в
полупространстве, приводится графическое сравнение профилей массовой
скорости Бозе-- и Ферми--газов. Выведена формула для значений
массовой скорости непосредственно у стенки. Приводится графическое
сравнение значений массовой скорости непосредственно у стенки
для Бозе-- и Ферми--газов как функций приведенного химического
потенциала, представляющего собой отношение химического
потенциала к произведению постоянной Больцмана на абсолютную
температуру.

Приведен численный анализ полученных результатов, представленный в виде
графиков на рис. 1--6. На рис. 2 приведено сравнение
коэффициентов изотермического скольжения для Бозе-- и
Ферми--газов. Различие между этими коэффициентами, как видно из
рис. 2, наблюдается в диапазоне $-4<\alpha<0$, при $\alpha<-4$
различие между коэффициентами несущественно. Точно такая же
ситуация наблюдается и для значений коэффициентов массовой
скорости непосредственно у стенки для Бозе-- и Ферми--газов (см. рис. 6).
На рис. 3, в отличие от рис. 2, добавлена кривая $3$, отвечающая
коэффициенту изотермического скольжения Бозе--газа с частотой
столкновений, пропорциональной модулю скорости молекул.

На рис. 4 приводится распределение массовой скорости в
полупространстве Бозе--газа. На рис. 5 приводится сравнение
распределений массовой скорости в полупространстве для Бозе-- и
Ферми--газов. Это сравнение показывает, что величина массовой
скорости больше для Бозе--газов, чем для Ферми--газов.

\end{document}